\documentclass[pdftex,12pt]{article}
\pdfoutput=1
\usepackage{graphicx}
\usepackage{txfonts}

\usepackage[colorlinks=true,citecolor=blue,breaklinks=true]{hyperref}
\usepackage{natbib}

\hypersetup{
pdftitle={Height distribution of the power of 3-min oscillations over sunspots},
pdfkeywords={Sun: sunspots -- Sun: oscillations -- Sun: chromosphere -- Sun: photosphere},
pdfauthor={D.Y. Kolobov}
}

\title{Height distribution of the power of 3-min oscillations over sunspots}
\author{N.I.\,Kobanov, D.Y.\,Kolobov, S.A.\,Chupin \\
\small Institute of Solar-Terrestrial Physics, \\
\small Russia, 664033, Irkutsk p/o box 291; Lermontov st., 126a  \\[-0.8ex]
\small kobanov@iszf.irk.ru \\
\and V.M.\,Nakariakov \\
\small Physics Department, University of Warwick, Coventry CV4 7AL, UK \\
\small V.Nakariakov@warwick.ac.uk
}

\date{\small{28 September 2010}, published in A\&A 525, A41 (2011)}

\begin{document}
\maketitle

\begin{abstract}
\par \textbf{Context.} The height structure of 3-min oscillations over sunspots
is studied in the context of the recently discovered effect of height inversion:
over the umbra, the spatial location of the maximum of chromospheric 3-min
oscillation power corresponds to the relative decrease in the power of
photospheric oscillations.

\par \textbf{Aims.} We investigate whether the height inversion of the power of
3-min oscillations is a common feature of  the spatial structure of the
oscillations for the majority of sunspots.

\par \textbf{Methods.} Spectrogram sequences of H$\alpha$~6563\AA\ and
Fe\,\textsc{i}~6569\AA\ over sunspots, acquired with very high cadency (about
2\,s or better) are obtained. The distribution of the oscillation power of the
line-of-sight velocity signal is studied by using methods of wavelet frequency
filtration and Fourier analysis.

\par \textbf{Results.} The effect of the height inversion is found in 9 of 11
analyzed active regions. The interpretation of this effect is possibly connected
to both the decrease in the level of photosphere in sunspot umbrae and the
magnetic field topology.
\end{abstract}

\section{Introduction}
\par Powerful 3-min oscillations of the line-of-sight (LOS) velocity, detected
in the Doppler shifts of the chromospheric spectral lines above sunspots have
been subject to intensive studies for several decades
\citep{beckers1969uf,giovanelli1972,zirin1972,lites1992a}. The interest in this
phenomenon is justified for several important reasons. First of all, it is
connected to the transfer of the energy of photospheric motions by these waves
to the corona. There is observational evidence of 3-min field-aligned
compressible waves in the corona over sunspots, usually seen in 171\AA\ and
195\AA\ passbands of TRACE (see, e.g. \citep{2006RSPTA.364..461D}), which may be
associated with the leakage of the 3-min sunspot oscillations. As the waves are
interpreted as slow magnetoacoustic waves that propagate almost parallel to the
magnetic field, their path highlights the magnetic connectivity of the
photospheric and coronal plasmas. In addition, these waves are a useful tool for
probing solar coronal plasmas in the quickly developing research field of
coronal seismology \citep{2003A&A...404L...1K, 2007ApJ...656..598W}. Moreover,
the relationship between 3-min oscillations in sunspots and the appearance of a
similar periodicity in flaring energy releases has been established
\citep{sych2009}. Another useful application of 3-min oscillations is in
attempting to reveal the internal, sub-photospheric structuring of sunspots
(e.g. \citep{2008SoPh..251..501Z}).

\par 3-min sunspot oscillations are presently studied in the optical band by
both filtering and spectral methods, as well as in radio and UV bands
\citep{gelfreikh1999,oshea2002a,rouppe2003,bloomfield2007a,nagashima2007,
jess2007, tziotziou2007a,balthasar2008}.
\citep{kobanov2004a,kobanov2006,bloomfield2007a} demonstrated that 3-min
oscillations in the umbral chromosphere are not the source of the running
penumbral waves and concluded that the apparently horizontally propagating waves
correspond to the ``visual pattern'' scenario.

\par Another identified peculiarity in the behaviour of 3-min oscillations has
been the effect of height inversion \citep{kobanov2008}: at the chromospheric
level, the spatial localisation of the maximum of 3-min oscillation power,
usually over the umbra, often corresponds to the minimum of these oscillations
at the photospheric level. One would expect that the spatial distribution of
3-min oscillations at both heights is similar, although at the photospheric
level the oscillation power in the umbra is lower than in the penumbra and
adjacent regions. This finding is consistent with the results of previous
studies. In particular,  it was pointed out that in the umbra, at the
photospheric height, 3-min oscillations are either not detected at all
\citep{balthasar1987} or of very low amplitude relative to the level of noise
\citep{lites1985a}. We note that the suppression of the photospheric oscillation
power in active regions has also been detected by the helioseismological methods
\citep{braun1990,braun1995,nicholas2004}. The analysis of the brightness
fluctuations observed in the G-band with Hinode/SOT detected the decrease in the
broadband oscillation power in central parts of sunspots \citep{nagashima2007}.

\par The aim of this paper is to perform a comparative analysis of 3-min
oscillations for a number of sunspots, demonstrate that the height inversion is
a statistically significant and reproducible effect, and contribute to its
interpretation. Our approach is based on the use of high cadence optical data,
with the time resolution of about 2~s, and upon the consideration of 3-min
oscillations in a narrow band, 4.7--6.7\,mHz. The paper is organised as follows.
In the next section, we describe the instrument used in the observations and
data reduction. In Section 3, we present the results obtained. The results
obtained are summarised in the conclusions.

\section{The instrument and data reduction}
\label{sec2}

\par
\begin{figure}[t]
  \centering
  \includegraphics[width=5cm]{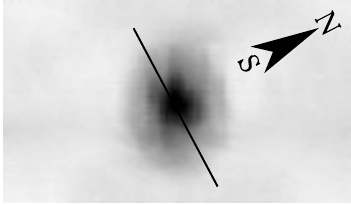}
  \caption{An example of the spectrograph slit orientation in the observations of the
  sunspot NOAA~791 (the scale of the slit is not kept).}
\label{fig:spot-slit}
\end{figure}

 We use data obtained with the Horizontal Solar Telescope at the Sayan Solar
Observatory, Russia \citep{2009ARep...53..957K}. The diameter of the mirror of
the instrument is 80~cm, hence the theoretically possible spatial resolution is
0.2$^{\prime\prime}$. However, because of the Earth's atmospheric conditions the
resolution is usually about 1$^{\prime\prime}$. The guiding system carries out
targeting and object capturing with a precision no worse than
1$^{\prime\prime}$. In observations, an astronomical CCD of the resolution
256$\times$1024 pixels was used. The spectral snapshot contains information
about a spatial region of the size 63$\times$0.5$^{\prime\prime}$, defined by
the slit of the spectrograph. For each spatial element, we obtain the spectrum
of the width of 8.5\AA\ (for the spectrograph dispersion of the fifth order in
the vicinity of the  H$\alpha$~6563\AA\ line). The image was rotated in a way to
position the spectrograph slit at the sunspot centre in either the north-south
or east-west directions (see \autoref{fig:spot-slit}), by using a Dove prism in
front of the spectrograph. The duration of observational sequences was about one
hour with a cadence time of about two seconds. The majority of the observational
sequences were recorded with the use of a deflector \citep{kobanov2001eit},
which provided us with information about the LOS velocity, spectral line
intensity, and the LOS component of the photospheric magnetic field (by the
Fe\,\textsc{i}~6569\AA\ line) simultaneously.

\par The information was recorded in a digital data cube of the following
structure. Two dimensions corresponded to the pixel number along the
spectrograph slit and dispersion (hence the spatial and spectral information),
while the third dimension contained the intensity of the emission at the certain
instant of time. A dedicated software package was then applied to reduce the
data cube to a two-dimensional array. The array contained the distribution of
the intensity and the LOS components of the velocity and magnetic field along
the slit at each instant of time. In the analysis of oscillatory processes, the
time sequences of each spatial pixel were studied by the Fourier and wavelet
techniques. The wavelet analysis was performed according to the methodology
described in \cite{torrence1998} with the 6th-order Morlet basis function.

\par The LOS component of the velocity was obtained by the lambdameter technique
\citep{rayrole1967}. In this method, the spectral position of a chosen line is
determined by two virtual slits. The distance between the slits is known and
remains constant during the measurements. Initially, the slits are situated at
equal distances from the line centre,  hence have equal intensities. If in the
next spectrogram the line is displaced, the intensities measured by the slits
differ. Displacing the slits to make their intensities equal allows one to
determine the new location of the line. Since this method is based upon the
determination of the relative position of the line, it is not affected by the
variation in the intensity in time, from one spectrogram to another.

\par In the processing of the observational sequences obtained with the
deflector, the LOS velocity is determined by the Doppler formula directly from
the displacement of H$\alpha$ line, as the splitting of the line by the
deflector is insignificant. In the case of another line used in this study,
Fe\,\textsc{i}, it is completely split by the deflector into the  $\sigma_{-}$
and $\sigma_{+}$ components. The LOS velocity is proportional to the sum of the
displacements of these components, while the magnetic field is proportional to
the difference in the displacements. The magnetic field is then calculated
according to the Zeeman effect for a simple triplet. Telluric line near
H$\alpha$ was used to eliminate the spectrograph noise.

\par A possible error in the determination of the line position may be related
to the asymmetry of the line shape. The asymmetry can be caused by blending of
the H$\alpha$ line, or by the variation in the LOS speeds between the heights
covered by the optically thick H$\alpha$ line. However, as in this study we
consider not the absolute values of the LOS speeds, but the frequencies of their
time variations, the latter effect not being likely to cause significant error.
In this paper, we did not discuss the results of the magnetic field
measurements.

\section{Results}
\label{sec3}

\par In our study, we analysed 11 active regions (AO) observed during
2002--2007. A list of the AO with their locations on the solar disk, the time of
the observation,  and results is presented in \autoref{table:AO}. The duration
of each sequence is about one hour, where $P_{p}$ and $P_{c}$ are the ratios of
the mean power of the 3 min oscillations in the spot umbra to the mean power in
the penumbra for the photospheric ($P_{p}$) and chromospheric ($P_{c}$) level
correspondingly.

\par
\begin{figure}[t]
  \centering
  \includegraphics[width=10cm]{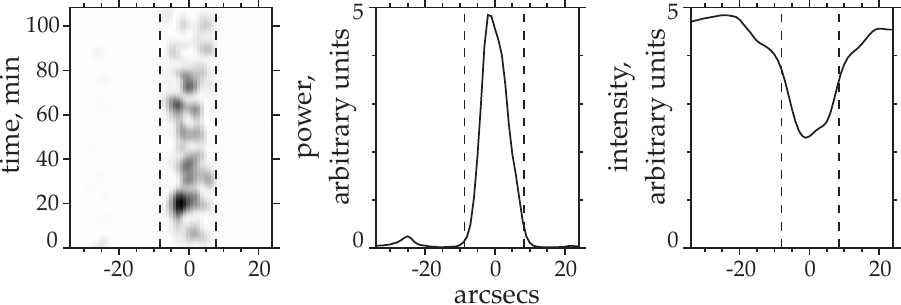}
  \caption{The distribution of the wavelet power of the LOS velocity in the range
  4.7--6.7\,mHz in the chromosphere of NOAA 810 observed in H$\alpha$. Left panel:
  time-spatial dynamics of the 3-min oscillations. Central panel: Profile of the time-averaged
  spatial distribution of the wavelet power. Right panel: the profile of the intensity observed in the
  continuum. The vertical dashed lines indicate the umbra-penumbra boundary.}
  \label{fig:NOAA810example}
\end{figure}

 The spatial localisation of the 3-min oscillations of the LOS velocity has
been studied using both Fourier and wavelet analyses. In both cases, the
analysed spectral interval was taken to be from 4.7 to 6.7~mHz.
\autoref{fig:NOAA810example} shows a typical wavelet spectrum that gives the
time-spatial dynamics of the 3-min oscillation, as well as the time-averaged
spatial distribution of the oscillation power.

\begin{table}
\caption{Observational sequences studied in this paper.}
\label{table:AO}
\centering
\begin{tabular}{c c c c c}
\hline\hline
NOAA & Solar coordinates & Date & Start time, UT & P$_{p}$/P$_{c}$\\
\hline
051     &S17E01&02.08.2002&00:45&0.5/12\\
051     &S16E10&01.08.2002&06:52&0.8/12.9\\
105 L   &S08E39&11.09.2002&01:21&0.6/9.2\\
613     &S09W03&20.05.2004&03:25&0.6/12.2\\
621     &S14W06&04.06.2004&05:43&---\\
657     &N10W03&13.08.2004&01:44&0.6/7.3\\
661 L   &N06E42&15.08.2004&23:57&0.7/3.7\\
661 L   &N07E20&18.08.2004&00:16&0.7/11.4\\
661 F   &N07E19&18.08.2004&01:01&{1.2/38.2}\\
791 L   &N13E21&26.07.2005&01:37&0.5/5.1\\
791 L   &N13E20&26.07.2005&03:27&{2.1/15.6}\\
791 L   &N13E08&27.07.2005&00:53&0.3/26.5\\
791 L   &N13E06&27.07.2005&01:26&0.5/16.3\\
791 L   &N13E06&27.07.2005&03:06&0.6/25.3\\
791 L   &N13E05&27.07.2005&04:46&0.4/33.3\\
794     &S11E32&04.08.2005&02:47&0.7/11\\
794     &S11E32&04.08.2005&03:38&1.1/18.6\\
794     &S11E05&05.08.2005&01:22&0.8/93\\
810     &N10E23&21.09.2005&02:45&0.9/9.2\\
810     &N10E23&21.09.2005&06:48&0.8/6.3\\
810     &N09W07&23.09.2005&07:15&{1.5/9.2}\\
886 L   &N07E24&25.05.2006&00:47&0.8/3.5\\
886 L   &N08E02&26.05.2006&04:25&0.7/4.7\\
963 L   &S07E62&08.07.2007&23:37&0.5/4\\
\hline
\end{tabular}
\end{table}

\begin{figure}[]
  \centering
  \includegraphics[width=12cm]{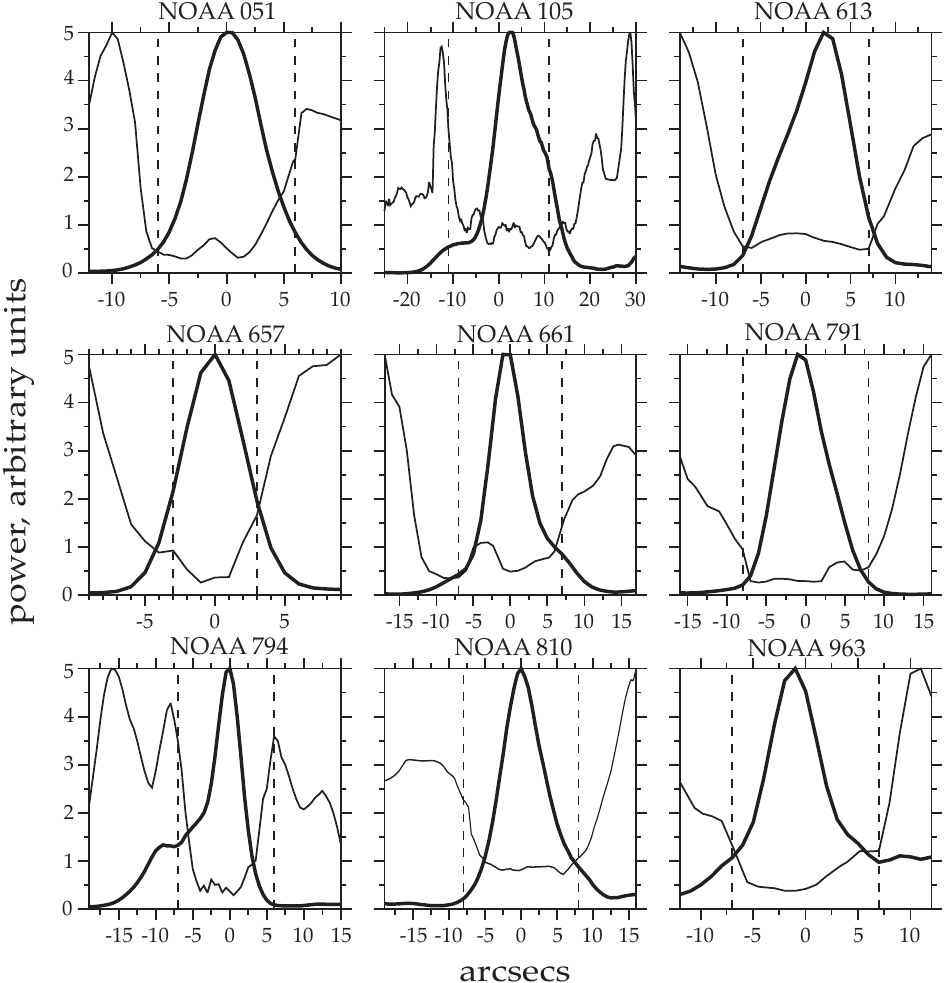}
  \caption{The effect of the height inversion for the power of 3-min
  oscillations  in  sunspot umbrae. The panels show time-averaged spatial profiles of
  3-min oscillations of the LOS velocity measured in different
  active regions. Thin lines correspond to the photospheric signal (Fe\,\textsc{i}), and the thick
  curves to the chromospheric signal (H\,$\alpha$). The vertical dashed lines indicate the
  umbra-penumbra boundary. }
  \label{fig:manyexamples}
\end{figure}
\begin{figure}[]
  \centering
  \includegraphics[width=12cm]{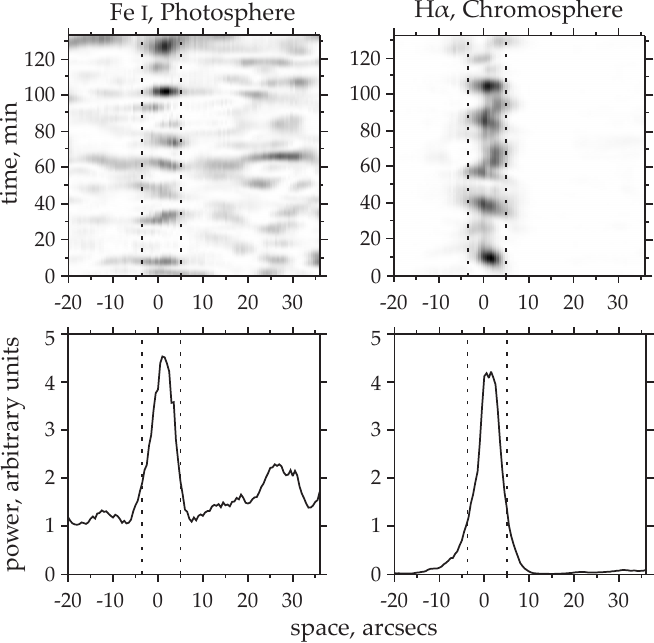}
  \caption{The absence of the height inversion effect for the leading
  sunspot of NOAA~791 in the observational sequences on 2005 July 26, 03:27--5:40\,UT.
In the same sunspot, the effect of the height inversion is presented in other five time series
(one of them is shown in \autoref{fig:manyexamples}).
  The figure shows the spatial distribution of the wavelet power of the LOS oscillations in the band
  4.7--6.7\,mHz.  \textit{Top panels:} Time-spatial dynamics of the wavelet spectra.
  \textit{Bottom panels}: Time-averaged profiles of the wavelet spectra.
  The vertical dashed lines indicate the umbra-penumbra boundary. }
  \label{fig:violator}
\end{figure}

\par  Our analysis demonstrated that at the chromospheric level in all
considered sunspots except the small sunspot NOAA\,621, the maxima of 3-min
oscillation power are situated near the centres of the umbrae. At the
photospheric level, in the most considered time series, these regions were found
to correspond to the suppression of the 3-min oscillation power (see
\autoref{fig:manyexamples}). The criterion used in the detection of the effect
of the height inversion is the occurrence of a local minimum in the spatial
distribution of the photospheric narrowband power and its absence at the
chromospheric level. The spatial coordinates in all plots of
\autoref{fig:manyexamples} were restricted by the external boundaries of the
penumbra. The effect can be quantified by the ratio of the mean power in the
umbra (averaged over the whole umbra) to the mean power in the penumbra
(averaged over the whole penumbra). In the chromosphere, this ratio was found to
be greater than one, while in the photosphere it is smaller than one (see
$P_{p}/P_{c}$, \autoref{table:AO}). We analysed 24 time series from 11 active
regions. In 18 series, we found the effect of height inversion, in 3 series the
situation was the opposite, and for 2 series we derived uncertain results (the
photospheric mean power ratio was 0.9--1.1). One time series of NOAA\,621 shows
the absence of the 3-min LOS velocity oscillations in the chromosphere.  We
should also point out that in certain time intervals in some sunspots the
maximum of 3-min oscillations at the photospheric level can be found in the
umbra. For example, for NOAA\,791, in six analysed time sequences, the effect of
the height inversion was found in five cases, while in one case the photospheric
maximum of the 3-min oscillations was localised in the umbra, hence exactly
coincident with the chromospheric maximum (see \autoref{fig:violator}). Wavelet
spectra of the photospheric and chromospheric signals are shown in the upper
panels. At first glance, the increases and decreases in the oscillation power at
those heights seem to occur simultaneously. However, the detailed phase analysis
discussed below does not support this impression.

\par We are thus inclined to conclude that in the majority of the observed cases
the spatial location of the maxima of 3-min chromospheric oscillations
correspond to the suppression of these oscillations at the photospheric level.
One possible interpretation of this effect may be the decrease in the
photospheric level observed in Fe\,\textsc{i}~6569\AA\ (the Wilson depression),
as it is known that the amplitude of acoustic oscillations decreases in deeper
regions  \citep{lites1992a,sigwarth1997}. The question is whether the difference
in heights, associated with the Wilson depression, is responsible for the
observed decrease in the oscillation amplitude by a factor of two.

\par According to \citet{1983SoPh...88...71B} and \cite{2009SoPh..260....5W},
the depression ($\tau$=1) of the umbral photosphere with respect to the penumbra
(the Wilson effect) is about 700 km.  This difference in height can lead to the
decrease in the 3-min oscillation amplitude by a factor of 2
\citep{1998ApJ...497..464L}. This decrease is consistent with the 3-min LOS
velocity power distribution at the photospheric level
(\autoref{fig:manyexamples}), at least within the order of the magnitude.

\citet{parnell1969} used the formation depth of the line Fe I 6569 to study the
granulation. They determined the upper limit to the line formation height to be
250 km. According to the broadly used atlas of the solar spectrum
\citep{moore1966}, this line is suppressed in sunspots, hence one can assume
that its optical thickness in the umbra is not higher. In the Fe I 6569 line,
the Wilson effect is as pronounced as in the continuum.

\par The observed effect may also be caused by the topology of the sunspot
magnetic field. If we assume that a horizontally extended source of 3-min
oscillations is situated below the photosphere, in the umbra, the oscillations
in the form of slow magnetoacoustic waves would be guided by the vertical
magnetic field lines upwards and reach the chromosphere. At the edge of the
umbra and in the penumbra, the magnetic field lines are inclined from the
vertical direction, preventing the free propagation of the short period waves
\citep{bogdan2006a}.  Thus, under the canopy 3-min waves are able to propagate
in the vertical direction up to the magnetic dome only. \citet{fin2004} pointed
out that the 7-mHz oscillations are reflected by the active region canopy. The
partial reflection of the waves from the magnetic dome can lead to the observed
accumulation of the wave energy at the photospheric level under the regions with
a horizontal magnetic field. This may explain the relative increase in the 3-min
photospheric oscillation power in the regions outside the umbra, and the
corresponding decrease in the chromospheric oscillation power in these  regions.

\par Moreover, it is necessary to consider a possible contribution of the \lq\lq
power halo\rq\rq \citep{1999ApJ...510..494L,1999ApJ...513L..79B}. Our
observations were focused on the sunspots, and, unfortunately, studying the
sunspots' vicinities is beyond the scope of the data for most time series.

\par
\begin{figure}[t]
  \centering
  \includegraphics[width=11.5cm]{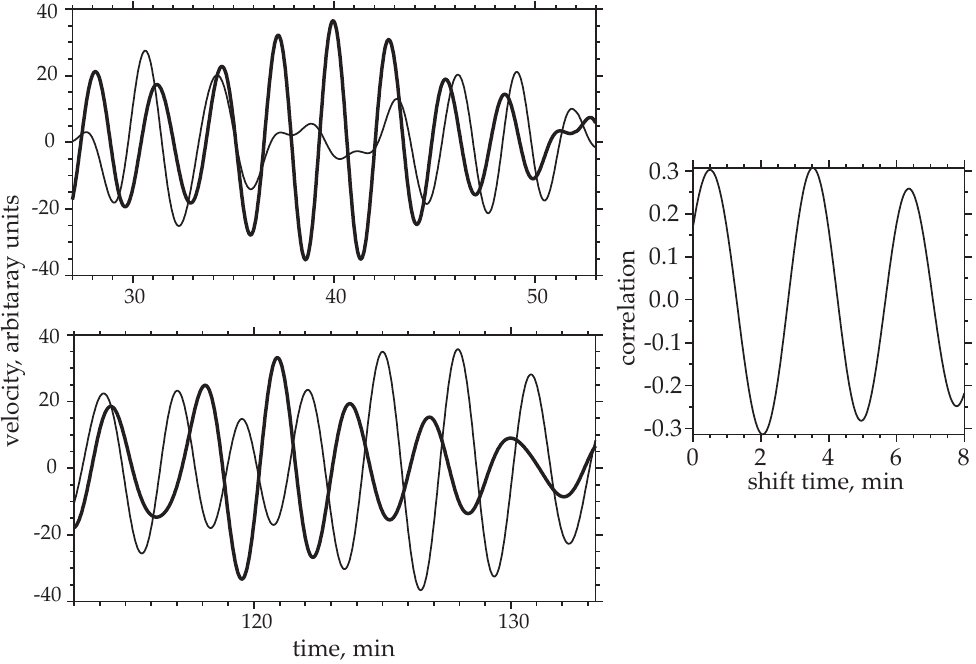}
  \caption{\textit{Left panel.} Comparison of phases of chromospheric (thick line) and photospheric (thin line)
  narrowband (4.7-6.7~mHz) LOS velocity signals in the centre of the umbra of NOAA 791
  (the leading spot) in different time intervals. \textit{Right panel.}
Correlation between the chromospheric and  photospheric signals of the LOS velocity for the whole time series.
  }
  \label{fig:velocityphasedelay}
\end{figure}

 In the majority of the observational sequences, the power of photospheric
3-min oscillations in the umbra was low, thus it was impossible to estimate the
phase delay between the photospheric and chromospheric oscillatory motions in
the umbra. However, the above-mentioned unusual observational sequences of NOAA
791, containing the umbral maximum of the photospheric 3-min oscillations,
allows us to carry out the analysis of the phase shifts between the signals at
different heights. Unfortunately, the direct comparison of the photospheric and
chromospheric signals does not provide us with any conclusive results.
\autoref{fig:velocityphasedelay} shows that there is no stable phase difference
between narrowband photospheric and chromospheric signals. In some time
intervals (e.g. 45--50 min), the chromospheric signal even precedes the
photospheric signal. There are also time intervals (e.g. 112--130~min) when the
time lag gradually increases from 20~s to 140~s. The correlation between the
signals for the whole time series is presented in the right panel of
\autoref{fig:velocityphasedelay}.

\par
\begin{figure}[t]
  \centering
  \includegraphics[width=11.5cm]{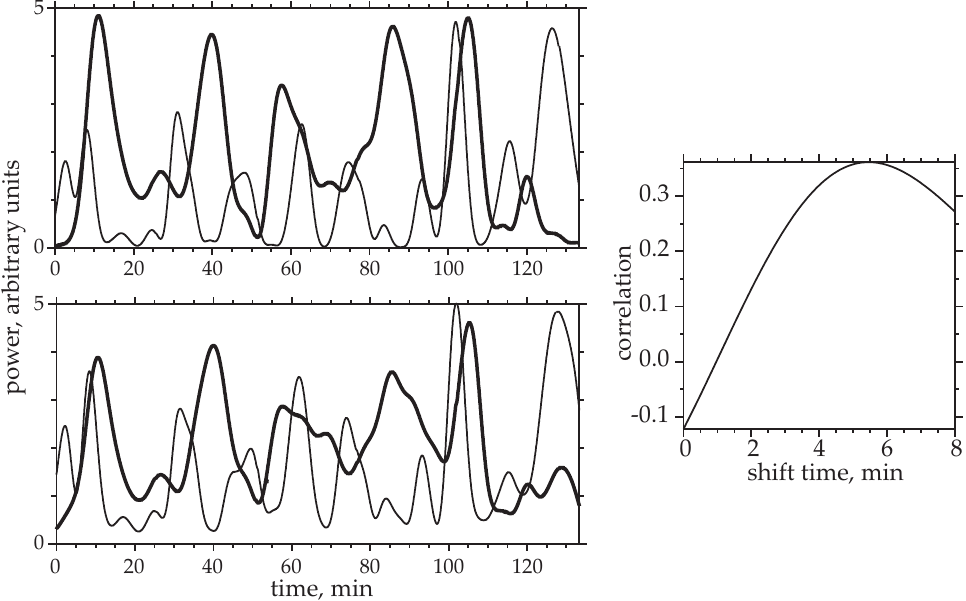}
  \caption{The time variation in the power of the
  chromospheric (thick line) and photospheric (thin line)
  narrowband (4.7-6.7~mHz) LOS velocity signals in the umbra of the leading spot of NOAA 791.
  The upper panel shows the signals of the spatial element of 1$^{\prime\prime}$ at the centre
  of the umbra.  The bottom panel shows the signals averaged over the 10$^{\prime\prime}$
  region in the umbra. Right panel shows the
  correlation between the signals presented in the upper panel.  }
  \label{fig:powerphasedelay}
\end{figure}

 A similar negative result comes from the comparison of the power maxima of
the photospheric and chromospheric signals (\autoref{fig:powerphasedelay}). The
detuning of the signals is so pronounced that one has the impression that the
photospheric and chromospheric signals are disconnected. Thus, the data obtained
does not allow us to measure the phase lag between the photospheric and
chromospheric 3-min oscillations directly, to derive the phase speed. However,
the average time lag between the appearance of the maxima of the oscillation
power at those two heights could be measured and is about 150~s. Taking the
distance between the heights of observations to be 2000~km
\citep{vernazza1981,white1966}, we obtain the vertical group speed of about
13\,$\mathrm{km\,s^{-1}}$.


On the other hand, taking into account the correlogram shown in the right panel of \autoref{fig:powerphasedelay}, the group speed can be estimated as 6\,km/s. This discrepancy is due to the ambiguity of the phase connection between the photospheric and chromospheric signals. Moreover, the observed correlation coefficient is rather low, about 0.35, and is observed only in one case. Thus, the present study does not allow us to make confident and statistically-significant conclusions about the wave connectivity at the photospheric and chromospheric levels, based upon the correlation analysis.
One possible reason for the apparent out-of-phase behaviour
of the photospheric and chromospheric oscillations may be the partial
reflection of the slow magnetoacoustic waves from the density gradient over the umbra,
which causes a complicated time-dependent superposition of the propagating and standing
waves in the photosphere.

\section{Conclusions}

\par We have analysed the spatial distribution of narrowband 3-min oscillations of the LOS
velocity at the photospheric and chromospheric levels in eleven sunspots. The observations
dedicated to this study were carried out with the
80~cm Horizontal Solar Telescope at the Sayan Solar Observatory in the Fe\,\textsc{i}~6569\AA\
and H$\alpha$~6563\AA\ lines, respectively. The spectral resolution of the instrument is 8.5\,m{\AA}.
The cadence time of the measurements was about 2~s.
At the chromospheric level in ten of the eleven considered sunspots,
the maxima of 3-min oscillation power were found to be situated
near the centres of the sunspot umbrae. In nine of the eleven sunspots at the photospheric level,
the 3-min oscillation power was found to be suppressed in the umbra relative to the
peripheral regions.

\par Thus, in the umbra, the chromospheric oscillations have higher power than outside,
while, in contrast, the photospheric oscillations in the umbra have systematically lower
amplitude than in the surrounding regions.
This effect may be termed a height inversion of 3-min oscillations in sunspots.
Phase shifts between narrowband 3-min amplitude photospheric and
chromospheric signals were found to vary in time, as well as between the maxima of their power.
The effect of height inversion may be attributed to the wave reflection caused by the inclined magnetic fields
in the peripheral regions of sunspots or to the Wilson depression of the photospheric
level in the umbra. However, detailed theoretical study of these phenomena is needed.

\small {
\section*{Acknowledgements}
The work is supported in part by the grant RFBR 08-02-91860-KO\_a, the Royal Society British-Russian
International Joint Project, grant 10-02-00153-a and grant of Federal Agency for Science and Innovation
(State Contract 02.740.11.0576). Wavelet software was provided by C.\,Torrence and G.\,Compo, and is available at
\href{http://paos.colorado.edu/research/wavelets}{http://paos.colorado.edu/\ research/wavelets}.
We would like to thank anonymous referee for the detailed consideration of our paper and for the constructive comments.
}
\clearpage

\bibliographystyle{spr-mp-sola-cnd}
\bibliography{kobanov-3mo}

\begin{thebibliography}{41}
\ifx \bisbn   \undefined \def \bisbn  #1{ISBN #1}\fi
\ifx \binits  \undefined \def \binits#1{#1}\fi
\ifx \bauthor  \undefined \def \bauthor#1{#1}\fi
\ifx \batitle  \undefined \def \batitle#1{#1}\fi
\ifx \bjtitle  \undefined \def \bjtitle#1{\textit{#1}}\fi
\ifx \bvolume  \undefined \def \bvolume#1{\textbf{#1}}\fi
\ifx \byear  \undefined \def \byear#1{#1}\fi
\ifx \bissue  \undefined \def \bissue#1{#1}\fi
\ifx \bfpage  \undefined \def \bfpage#1{#1}\fi
\ifx \blpage  \undefined \def \blpage #1{#1}\fi
\ifx \burl  \undefined \def \burl#1{\textsf{#1}}\fi
\ifx \href  \undefined \def \href#1#2{\textsf{#2}}\fi
\ifx \doiurl  \undefined \def
  \doiurl#1{\href{http://dx.doi.org/#1}{\textsf{#1}}}\fi
\ifx \betal  \undefined \def \betal{\textit{et al.}}\fi
\ifx \binstitute  \undefined \def \binstitute#1{#1}\fi
\ifx \bctitle  \undefined \def \bctitle#1{#1}\fi
\ifx \beditor  \undefined \def \beditor#1{#1}\fi
\ifx \bpublisher  \undefined \def \bpublisher#1{#1}\fi
\ifx \bbtitle  \undefined \def \bbtitle#1{\textit{#1}}\fi
\ifx \bedition  \undefined \def \bedition#1{#1}\fi
\ifx \bseriesno  \undefined \def \bseriesno#1{\textbf{#1}}\fi
\ifx \blocation  \undefined \def \blocation#1{#1}\fi
\ifx \bsertitle  \undefined \def \bsertitle#1{\textit{#1}}\fi
\ifx \bsnm \undefined \def \bsnm#1{#1}\fi
\ifx \bsuffix \undefined \def \bsuffix#1{#1}\fi
\ifx \bparticle \undefined \def \bparticle#1{#1}\fi
\ifx \barticle \undefined \def \barticle#1{}\fi
\ifx \botherref \undefined \def \botherref#1{}\fi
\ifx \url \undefined \def \url#1{\textsf{#1}}\fi
\ifx \bchapter \undefined \def \bchapter#1{}\fi
\ifx \bbook \undefined \def \bbook#1{}\fi
\ifx \bcomment \undefined \def \bcomment#1{#1}\fi
\ifx \oauthor \undefined \def \oauthor#1{#1}\fi
\ifx \citeauthoryear \undefined \def \citeauthoryear#1{#1}\fi
\def \endbibitem {}
\ifx \bconflocation  \undefined \def \bconflocation#1{#1} \fi

\bibitem[\protect\citeauthoryear{{Balthasar} and
  {Schleicher}}{2008}]{balthasar2008}
\begin{barticle}
\bauthor{\bsnm{{Balthasar}}, \binits{H.}},
\bauthor{\bsnm{{Schleicher}}, \binits{H.}}:
\byear{2008},
\bjtitle{\aap}
\bvolume{481},
\bfpage{811}.
doi:\doiurl{10.1051/0004-6361:20078279}.
\end{barticle}
\endbibitem

\bibitem[\protect\citeauthoryear{{Balthasar} and
  {Woehl}}{1983}]{1983SoPh...88...71B}
\begin{barticle}
\bauthor{\bsnm{{Balthasar}}, \binits{H.}},
\bauthor{\bsnm{{Woehl}}, \binits{H.}}:
\byear{1983},
\bjtitle{\solphys}
\bvolume{88},
\bfpage{71}.
doi:\doiurl{10.1007/BF00196178}.
\end{barticle}
\endbibitem

\bibitem[\protect\citeauthoryear{{Balthasar}, {Kueveler}, and
  {Wiehr}}{1987}]{balthasar1987}
\begin{barticle}
\bauthor{\bsnm{{Balthasar}}, \binits{H.}},
\bauthor{\bsnm{{Kueveler}}, \binits{G.}},
\bauthor{\bsnm{{Wiehr}}, \binits{E.}}:
\byear{1987},
\bjtitle{\solphys}
\bvolume{112},
\bfpage{37}.
\end{barticle}
\endbibitem

\bibitem[\protect\citeauthoryear{{Beckers} and {Tallant}}{1969}]{beckers1969uf}
\begin{barticle}
\bauthor{\bsnm{{Beckers}}, \binits{J.M.}},
\bauthor{\bsnm{{Tallant}}, \binits{P.E.}}:
\byear{1969},
\bjtitle{\solphys}
\bvolume{7},
\bfpage{351}.
\end{barticle}
\endbibitem

\bibitem[\protect\citeauthoryear{{Bloomfield}, {Lagg}, and
  {Solanki}}{2007}]{bloomfield2007a}
\begin{barticle}
\bauthor{\bsnm{{Bloomfield}}, \binits{D.S.}},
\bauthor{\bsnm{{Lagg}}, \binits{A.}},
\bauthor{\bsnm{{Solanki}}, \binits{S.K.}}:
\byear{2007},
\bjtitle{\apj}
\bvolume{671},
\bfpage{1005}.
doi:\doiurl{10.1086/523266}.
\end{barticle}
\endbibitem

\bibitem[\protect\citeauthoryear{{Bogdan} and {Judge}}{2006}]{bogdan2006a}
\begin{barticle}
\bauthor{\bsnm{{Bogdan}}, \binits{T.J.}},
\bauthor{\bsnm{{Judge}}, \binits{P.G.}}:
\byear{2006},
\bjtitle{Royal Society of London Philosophical Transactions Series A}
\bvolume{364},
\bfpage{313}.
\end{barticle}
\endbibitem

\bibitem[\protect\citeauthoryear{{Braun}}{1995}]{braun1995}
\begin{barticle}
\bauthor{\bsnm{{Braun}}, \binits{D.C.}}:
\byear{1995},
\bjtitle{\apj}
\bvolume{451},
\bfpage{859}.
doi:\doiurl{10.1086/176272}.
\end{barticle}
\endbibitem

\bibitem[\protect\citeauthoryear{{Braun} and {Duvall}}{1990}]{braun1990}
\begin{barticle}
\bauthor{\bsnm{{Braun}}, \binits{D.C.}},
\bauthor{\bsnm{{Duvall}}, \binits{T.L.} \bsuffix{Jr.}}:
\byear{1990},
\bjtitle{\solphys}
\bvolume{129},
\bfpage{83}.
doi:\doiurl{10.1007/BF00154366}.
\end{barticle}
\endbibitem

\bibitem[\protect\citeauthoryear{{Braun} and
  {Lindsey}}{1999}]{1999ApJ...513L..79B}
\begin{barticle}
\bauthor{\bsnm{{Braun}}, \binits{D.C.}},
\bauthor{\bsnm{{Lindsey}}, \binits{C.}}:
\byear{1999},
\bjtitle{\apjl}
\bvolume{513},
\bfpage{L79}.
doi:\doiurl{10.1086/311897}.
\end{barticle}
\endbibitem

\bibitem[\protect\citeauthoryear{{De Moortel}}{2006}]{2006RSPTA.364..461D}
\begin{barticle}
\bauthor{\bsnm{{De Moortel}}, \binits{I.}}:
\byear{2006},
\bjtitle{Royal Society of London Philosophical Transactions Series A}
\bvolume{364},
\bfpage{461}.
\end{barticle}
\endbibitem

\bibitem[\protect\citeauthoryear{{Finsterle} \textit{et~al.}}{2004}]{fin2004}
\begin{barticle}
\bauthor{\bsnm{{Finsterle}}, \binits{W.}},
\bauthor{\bsnm{{Jefferies}}, \binits{S.M.}},
\bauthor{\bsnm{{Cacciani}}, \binits{A.}},
\bauthor{\bsnm{{Rapex}}, \binits{P.}},
\bauthor{\bsnm{{McIntosh}}, \binits{S.W.}}:
\byear{2004},
\bjtitle{\apjl}
\bvolume{613},
\bfpage{L185}.
doi:\doiurl{10.1086/424996}.
\end{barticle}
\endbibitem

\bibitem[\protect\citeauthoryear{{Gelfreikh}
  \textit{et~al.}}{1999}]{gelfreikh1999}
\begin{barticle}
\bauthor{\bsnm{{Gelfreikh}}, \binits{G.B.}},
\bauthor{\bsnm{{Grechnev}}, \binits{V.}},
\bauthor{\bsnm{{Kosugi}}, \binits{T.}},
\bauthor{\bsnm{{Shibasaki}}, \binits{K.}}:
\byear{1999},
\bjtitle{\solphys}
\bvolume{185},
\bfpage{177}.
\end{barticle}
\endbibitem

\bibitem[\protect\citeauthoryear{{Giovanelli}}{1972}]{giovanelli1972}
\begin{barticle}
\bauthor{\bsnm{{Giovanelli}}, \binits{R.G.}}:
\byear{1972},
\bjtitle{\solphys}
\bvolume{27},
\bfpage{71}.
\end{barticle}
\endbibitem

\bibitem[\protect\citeauthoryear{{Jess} \textit{et~al.}}{2007}]{jess2007}
\begin{barticle}
\bauthor{\bsnm{{Jess}}, \binits{D.B.}},
\bauthor{\bsnm{{Andi{\'c}}}, \binits{A.}},
\bauthor{\bsnm{{Mathioudakis}}, \binits{M.}},
\bauthor{\bsnm{{Bloomfield}}, \binits{D.S.}},
\bauthor{\bsnm{{Keenan}}, \binits{F.P.}}:
\byear{2007},
\bjtitle{\aap}
\bvolume{473},
\bfpage{943}.
doi:\doiurl{10.1051/0004-6361:20077142}.
\end{barticle}
\endbibitem

\bibitem[\protect\citeauthoryear{{King}
  \textit{et~al.}}{2003}]{2003A&A...404L...1K}
\begin{barticle}
\bauthor{\bsnm{{King}}, \binits{D.B.}},
\bauthor{\bsnm{{Nakariakov}}, \binits{V.M.}},
\bauthor{\bsnm{{Deluca}}, \binits{E.E.}},
\bauthor{\bsnm{{Golub}}, \binits{L.}},
\bauthor{\bsnm{{McClements}}, \binits{K.G.}}:
\byear{2003},
\bjtitle{\aap}
\bvolume{404},
\bfpage{L1}.
doi:\doiurl{10.1051/0004-6361:20030763}.
\end{barticle}
\endbibitem

\bibitem[\protect\citeauthoryear{{Kobanov}}{2001}]{kobanov2001eit}
\begin{barticle}
\bauthor{\bsnm{{Kobanov}}, \binits{N.I.}}:
\byear{2001},
\bjtitle{Instruments and Experimental Techniques, Vol.~4, p.~110-115}
\bvolume{4},
\bfpage{110}.
\end{barticle}
\endbibitem

\bibitem[\protect\citeauthoryear{{Kobanov} and
  {Makarchik}}{2004}]{kobanov2004a}
\begin{barticle}
\bauthor{\bsnm{{Kobanov}}, \binits{N.I.}},
\bauthor{\bsnm{{Makarchik}}, \binits{D.V.}}:
\byear{2004},
\bjtitle{\aap}
\bvolume{424},
\bfpage{671}.
doi:\doiurl{10.1051/0004-6361:20035960}.
\end{barticle}
\endbibitem

\bibitem[\protect\citeauthoryear{{Kobanov}, {Kolobov}, and
  {Chupin}}{2008}]{kobanov2008}
\begin{barticle}
\bauthor{\bsnm{{Kobanov}}, \binits{N.I.}},
\bauthor{\bsnm{{Kolobov}}, \binits{D.Y.}},
\bauthor{\bsnm{{Chupin}}, \binits{S.A.}}:
\byear{2008},
\bjtitle{Astronomy Letters}
\bvolume{34},
\bfpage{133}.
doi:\doiurl{10.1007/s11443-008-2006-5}.
\end{barticle}
\endbibitem

\bibitem[\protect\citeauthoryear{{Kobanov}, {Kolobov}, and
  {Makarchik}}{2006}]{kobanov2006}
\begin{barticle}
\bauthor{\bsnm{{Kobanov}}, \binits{N.I.}},
\bauthor{\bsnm{{Kolobov}}, \binits{D.Y.}},
\bauthor{\bsnm{{Makarchik}}, \binits{D.V.}}:
\byear{2006},
\bjtitle{\solphys}
\bvolume{238},
\bfpage{231}.
doi:\doiurl{10.1007/s11207-006-0160-z}.
\end{barticle}
\endbibitem

\bibitem[\protect\citeauthoryear{{Kobanov}
  \textit{et~al.}}{2009}]{2009ARep...53..957K}
\begin{barticle}
\bauthor{\bsnm{{Kobanov}}, \binits{N.I.}},
\bauthor{\bsnm{{Kolobov}}, \binits{D.Y.}},
\bauthor{\bsnm{{Sklyar}}, \binits{A.A.}},
\bauthor{\bsnm{{Chupin}}, \binits{S.A.}},
\bauthor{\bsnm{{Pulyaev}}, \binits{V.A.}}:
\byear{2009},
\bjtitle{Astronomy Reports}
\bvolume{53},
\bfpage{957}.
doi:\doiurl{10.1134/S1063772909100072}.
\end{barticle}
\endbibitem

\bibitem[\protect\citeauthoryear{{Lindsey} and
  {Braun}}{1999}]{1999ApJ...510..494L}
\begin{barticle}
\bauthor{\bsnm{{Lindsey}}, \binits{C.}},
\bauthor{\bsnm{{Braun}}, \binits{D.C.}}:
\byear{1999},
\bjtitle{\apj}
\bvolume{510},
\bfpage{494}.
doi:\doiurl{10.1086/306560}.
\end{barticle}
\endbibitem

\bibitem[\protect\citeauthoryear{{Lites}}{1992}]{lites1992a}
\begin{bchapter}
\bauthor{\bsnm{{Lites}}, \binits{B.W.}}:
\byear{1992},
In: \beditor{\bsnm{{Thomas}}, \binits{J.H.}},
\beditor{\bsnm{{Weiss}}, \binits{N.O.}} (eds.)
\bbtitle{NATO ASIC Proc. 375: Sunspots. Theory and Observations},
\bfpage{261}.
\end{bchapter}
\endbibitem

\bibitem[\protect\citeauthoryear{{Lites} and {Thomas}}{1985}]{lites1985a}
\begin{barticle}
\bauthor{\bsnm{{Lites}}, \binits{B.W.}},
\bauthor{\bsnm{{Thomas}}, \binits{J.H.}}:
\byear{1985},
\bjtitle{\apj}
\bvolume{294},
\bfpage{682}.
doi:\doiurl{10.1086/163338}.
\end{barticle}
\endbibitem

\bibitem[\protect\citeauthoryear{{Lites}
  \textit{et~al.}}{1998}]{1998ApJ...497..464L}
\begin{barticle}
\bauthor{\bsnm{{Lites}}, \binits{B.W.}},
\bauthor{\bsnm{{Thomas}}, \binits{J.H.}},
\bauthor{\bsnm{{Bogdan}}, \binits{T.J.}},
\bauthor{\bsnm{{Cally}}, \binits{P.S.}}:
\byear{1998},
\bjtitle{\apj}
\bvolume{497},
\bfpage{464}.
doi:\doiurl{10.1086/305451}.
\end{barticle}
\endbibitem

\bibitem[\protect\citeauthoryear{{Moore}, {Minnaert}, and
  {Houtgast}}{1966}]{moore1966}
\begin{bbook}
\bauthor{\bsnm{{Moore}}, \binits{C.E.}},
\bauthor{\bsnm{{Minnaert}}, \binits{M.G.J.}},
\bauthor{\bsnm{{Houtgast}}, \binits{J.}}:
\byear{1966},
\bbtitle{{The solar spectrum 2935 A to 8770 A}}.
\end{bbook}
\endbibitem

\bibitem[\protect\citeauthoryear{{Nagashima}
  \textit{et~al.}}{2007}]{nagashima2007}
\begin{barticle}
\bauthor{\bsnm{{Nagashima}}, \binits{K.}},
\bauthor{\bsnm{{Sekii}}, \binits{T.}},
\bauthor{\bsnm{{Kosovichev}}, \binits{A.G.}},
\bauthor{\bsnm{{Shibahashi}}, \binits{H.}},
\bauthor{\bsnm{{Tsuneta}}, \binits{S.}},
\bauthor{\bsnm{{Ichimoto}}, \binits{K.}},
\bauthor{\bsnm{{Katsukawa}}, \binits{Y.}},
\bauthor{\bsnm{{Lites}}, \binits{B.}},
\bauthor{\bsnm{{Nagata}}, \binits{S.}},
\bauthor{\bsnm{{Shimizu}}, \binits{T.}},
\bauthor{\bsnm{{Shine}}, \binits{R.A.}},
\bauthor{\bsnm{{Suematsu}}, \binits{Y.}},
\bauthor{\bsnm{{Tarbell}}, \binits{T.D.}},
\bauthor{\bsnm{{Title}}, \binits{A.M.}}:
\byear{2007},
\bjtitle{\pasj}
\bvolume{59},
\bfpage{631}.
\end{barticle}
\endbibitem

\bibitem[\protect\citeauthoryear{{Nicholas}, {Thompson}, and
  {Rajaguru}}{2004}]{nicholas2004}
\begin{barticle}
\bauthor{\bsnm{{Nicholas}}, \binits{C.J.}},
\bauthor{\bsnm{{Thompson}}, \binits{M.J.}},
\bauthor{\bsnm{{Rajaguru}}, \binits{S.P.}}:
\byear{2004},
\bjtitle{\solphys}
\bvolume{225},
\bfpage{213}.
doi:\doiurl{10.1007/s11207-004-4586-x}.
\end{barticle}
\endbibitem

\bibitem[\protect\citeauthoryear{{O'Shea}, {Muglach}, and
  {Fleck}}{2002}]{oshea2002a}
\begin{barticle}
\bauthor{\bsnm{{O'Shea}}, \binits{E.}},
\bauthor{\bsnm{{Muglach}}, \binits{K.}},
\bauthor{\bsnm{{Fleck}}, \binits{B.}}:
\byear{2002},
\bjtitle{\aap}
\bvolume{387},
\bfpage{642}.
doi:\doiurl{10.1051/0004-6361:20020375}.
\end{barticle}
\endbibitem

\bibitem[\protect\citeauthoryear{{Parnell} and {Beckers}}{1969}]{parnell1969}
\begin{barticle}
\bauthor{\bsnm{{Parnell}}, \binits{R.L.}},
\bauthor{\bsnm{{Beckers}}, \binits{J.M.}}:
\byear{1969},
\bjtitle{\solphys}
\bvolume{9},
\bfpage{35}.
doi:\doiurl{10.1007/BF00145725}.
\end{barticle}
\endbibitem

\bibitem[\protect\citeauthoryear{{Rayrole}}{1967}]{rayrole1967}
\begin{barticle}
\bauthor{\bsnm{{Rayrole}}, \binits{J.}}:
\byear{1967},
\bjtitle{Annales d'Astrophysique}
\bvolume{30},
\bfpage{257}.
\end{barticle}
\endbibitem

\bibitem[\protect\citeauthoryear{{Rouppe van der Voort}
  \textit{et~al.}}{2003}]{rouppe2003}
\begin{barticle}
\bauthor{\bsnm{{Rouppe van der Voort}}, \binits{L.H.M.}},
\bauthor{\bsnm{{Rutten}}, \binits{R.J.}},
\bauthor{\bsnm{{S{\"u}tterlin}}, \binits{P.}},
\bauthor{\bsnm{{Sloover}}, \binits{P.J.}},
\bauthor{\bsnm{{Krijger}}, \binits{J.M.}}:
\byear{2003},
\bjtitle{\aap}
\bvolume{403},
\bfpage{277}.
doi:\doiurl{10.1051/0004-6361:20030237}.
\end{barticle}
\endbibitem

\bibitem[\protect\citeauthoryear{{Sigwarth} and {Mattig}}{1997}]{sigwarth1997}
\begin{barticle}
\bauthor{\bsnm{{Sigwarth}}, \binits{M.}},
\bauthor{\bsnm{{Mattig}}, \binits{W.}}:
\byear{1997},
\bjtitle{\aap}
\bvolume{324},
\bfpage{743}.
\end{barticle}
\endbibitem

\bibitem[\protect\citeauthoryear{{Sych} \textit{et~al.}}{2009}]{sych2009}
\begin{barticle}
\bauthor{\bsnm{{Sych}}, \binits{R.}},
\bauthor{\bsnm{{Nakariakov}}, \binits{V.M.}},
\bauthor{\bsnm{{ Karlicky}}, \binits{M.}},
\bauthor{\bsnm{{Anfinogentov}}, \binits{S.}}:
\byear{2009},
\bjtitle{\aap}
\bvolume{505},
\bfpage{791}.
doi:\doiurl{10.1051/0004-6361/200912132}.
\end{barticle}
\endbibitem

\bibitem[\protect\citeauthoryear{{Torrence} and {Compo}}{1998}]{torrence1998}
\begin{barticle}
\bauthor{\bsnm{{Torrence}}, \binits{C.}},
\bauthor{\bsnm{{Compo}}, \binits{G.P.}}:
\byear{1998},
\bjtitle{Bulletin of the American Meteorological Society, vol.~79, Issue 1,
  pp.61-78}
\bvolume{79},
\bfpage{61}.
\end{barticle}
\endbibitem

\bibitem[\protect\citeauthoryear{{Tziotziou}
  \textit{et~al.}}{2007}]{tziotziou2007a}
\begin{barticle}
\bauthor{\bsnm{{Tziotziou}}, \binits{K.}},
\bauthor{\bsnm{{Tsiropoula}}, \binits{G.}},
\bauthor{\bsnm{{Mein}}, \binits{N.}},
\bauthor{\bsnm{{Mein}}, \binits{P.}}:
\byear{2007},
\bjtitle{\aap}
\bvolume{463},
\bfpage{1153}.
doi:\doiurl{10.1051/0004-6361:20066412}.
\end{barticle}
\endbibitem

\bibitem[\protect\citeauthoryear{{Vernazza}, {Avrett}, and
  {Loeser}}{1981}]{vernazza1981}
\begin{barticle}
\bauthor{\bsnm{{Vernazza}}, \binits{J.E.}},
\bauthor{\bsnm{{Avrett}}, \binits{E.H.}},
\bauthor{\bsnm{{Loeser}}, \binits{R.}}:
\byear{1981},
\bjtitle{\apjs}
\bvolume{45},
\bfpage{635}.
doi:\doiurl{10.1086/190731}.
\end{barticle}
\endbibitem

\bibitem[\protect\citeauthoryear{{Wang}, {Innes}, and
  {Qiu}}{2007}]{2007ApJ...656..598W}
\begin{barticle}
\bauthor{\bsnm{{Wang}}, \binits{T.}},
\bauthor{\bsnm{{Innes}}, \binits{D.E.}},
\bauthor{\bsnm{{Qiu}}, \binits{J.}}:
\byear{2007},
\bjtitle{\apj}
\bvolume{656},
\bfpage{598}.
doi:\doiurl{10.1086/510424}.
\end{barticle}
\endbibitem

\bibitem[\protect\citeauthoryear{{Watson}
  \textit{et~al.}}{2009}]{2009SoPh..260....5W}
\begin{barticle}
\bauthor{\bsnm{{Watson}}, \binits{F.}},
\bauthor{\bsnm{{Fletcher}}, \binits{L.}},
\bauthor{\bsnm{{Dalla}}, \binits{S.}},
\bauthor{\bsnm{{Marshall}}, \binits{S.}}:
\byear{2009},
\bjtitle{\solphys}
\bvolume{260},
\bfpage{5}.
doi:\doiurl{10.1007/s11207-009-9420-z}.
\end{barticle}
\endbibitem

\bibitem[\protect\citeauthoryear{{White} and {Wilson}}{1966}]{white1966}
\begin{barticle}
\bauthor{\bsnm{{White}}, \binits{O.R.}},
\bauthor{\bsnm{{Wilson}}, \binits{P.R.}}:
\byear{1966},
\bjtitle{\apj}
\bvolume{146},
\bfpage{250}.
\end{barticle}
\endbibitem

\bibitem[\protect\citeauthoryear{{Zhugzhda}}{2008}]{2008SoPh..251..501Z}
\begin{barticle}
\bauthor{\bsnm{{Zhugzhda}}, \binits{Y.D.}}:
\byear{2008},
\bjtitle{\solphys}
\bvolume{251},
\bfpage{501}.
doi:\doiurl{10.1007/s11207-008-9251-3}.
\end{barticle}
\endbibitem

\bibitem[\protect\citeauthoryear{{Zirin} and {Stein}}{1972}]{zirin1972}
\begin{barticle}
\bauthor{\bsnm{{Zirin}}, \binits{H.}},
\bauthor{\bsnm{{Stein}}, \binits{A.}}:
\byear{1972},
\bjtitle{\apjl}
\bvolume{178},
\bfpage{L85}.
\end{barticle}
\endbibitem

\end{thebibliography}

\end{document}